\documentclass[10pt, twocolumn]{IEEEtran}
\usepackage{hyperref}
\usepackage{booktabs}
\usepackage{graphicx}
\usepackage{epstopdf}

\usepackage{cite}

\ifCLASSINFOpdf
\else
\fi
\usepackage{amsmath}
\usepackage{algorithmic}

\usepackage{array}
\usepackage{color}

\begin{document}
%
\title{Fast On-orbit Pulse Phase Estimation of X-ray Crab Pulsar for XNAV Flight Experiments}


\author{\IEEEauthorblockN{Yidi Wang\IEEEauthorrefmark{1},
Shuangnan Zhang\IEEEauthorrefmark{2}\IEEEauthorrefmark{3},
Minyu Ge\IEEEauthorrefmark{2}, 
Wei Zheng\IEEEauthorrefmark{1},
Xiaoqian Chen\IEEEauthorrefmark{1},
Shijie Zheng\IEEEauthorrefmark{2},
Fangju Lu\IEEEauthorrefmark{2}}\\
\IEEEauthorblockA{\IEEEauthorrefmark{1}College of Aerospace Science and Engineering, National University of Defense Technology, Changsha 410073, China}\\
\IEEEauthorblockA{\IEEEauthorrefmark{2}Key Laboratory of Particle Astrophysics, Institute of High Energy Physics, Chinese Academy of Science, Beijing 100049, China}\\
\IEEEauthorblockA{\IEEEauthorrefmark{3}University of Chinese Academy of Sciences, Chinese Academy of Science, Beijing 100049, China}\\
\thanks{Manuscript received December 1, 2012; revised August 26, 2015. 
Corresponding author: Yidi Wang (email: wangyidi$\_$nav@163.com), Shuangnan Zhang (email: zhangsn@ihep.ac.cn.), Wei Zheng (email: zhengwei@nudt.edut.cn).}}

\markboth{Journal of \LaTeX\ Class Files,~Vol.~14, No.~8, August~2015}%
{Yidi Wang \MakeLowercase{\textit{et al.}}: X-ray Pulsar Navigation using On-orbit Pulsar Timing}
%



\IEEEtitleabstractindextext{%
\begin{abstract}
  The recent flight experiments with  Neutron Star Interior Composition Explorer (\textit{NICER}) and \textit{Insight}-Hard X-ray Modulation Telescope (\textit{Insight}-HXMT) have demonstrated the feasibility of X-ray pulsar-based navigation (XNAV) in the space. However, the current pulse phase estimation and navigation methods employed in the above flight experiments are computationally too expensive for handling the Crab pulsar data. To solve this problem, this paper proposes a fast algorithm of on-orbit estimating the pulse phase of Crab pulsar called X-ray pulsar navigaTion usIng on-orbiT pulsAr timiNg (XTITAN). The pulse phase propagation model for Crab pulsar data from \textit{Insight}-HXMT and \textit{NICER} are derived. When an exposure on the Crab pulsar is divided into several sub-exposures, we derive an on-orbit timing method to estimate the hyperparameters of the pulse phase propagation model. Moreover, XTITAN is improved by iteratively estimating the pulse phase and the position and velocity of satellite. When applied to the Crab pulsar data from \textit{NICER}, XTITAN is 58 times faster than the grid search method employed by \textit{NICER} experiment. When applied to the Crab pulsar data from \textit{Insight}-HXMT, XTITAN is 180 times faster than the Significance Enhancement of Pulse-profile with Orbit-dynamics (SEPO) which was employed in the flight experiments with \textit{Insight}-HXMT. Thus, XTITAN is computationally much efficient and has the potential to be employed for onboard computation.
\end{abstract}

\begin{IEEEkeywords}
Pulsar Navigation, Pulsar Signal Processing, Spacecraft Autonomous Navigation, Deep Space Exploration
\end{IEEEkeywords}}

\maketitle

\IEEEdisplaynontitleabstractindextext

%
\IEEEpeerreviewmaketitle

\section{Introduction}
\label{sect:intro}
%
%
%
%
\IEEEPARstart{W}{hen} the footprints of human go further into the deep space, the current ground-based tracking system cannot afford a timely and effective support because the distance between the spacecraft and the Earth dramatically grows. Thus, an autonomous navigation system is urgently needed. The image-based autonomous navigation system has been already applied to deep space explorations, but its positioning performance will degrade when there are no planets nearby \cite{Liu_2019}. In this case, the X-ray pulsar-based navigation (XNAV) is a promising solution. XNAV was first introduced in the 1980s, and its theoretical framework been gradually developed through the next 40 years \cite{Sheikh_2006, Wang_2014, Gui2019,Runnels2021}. However, most of the previous literatures concerning XNAV were based on simulations. How XNAV performs via real pulsar data was an open problem until the United States performed the XNAV onboard demonstration with the Neutron Star Interior Composition Explorer (\textit{NICER}) on the International Space Station (ISS) in 2018 \cite{Mitchell_2018,Winternitz2018}. China also verified the orbit determination performance of XNAV with the Crab pulsar data from \textit{Insight}-Hard X-ray Modulation Telescope (\textit{Insight}-HXMT) in 2019 \cite{Zheng_2019,Yusong2021}.

During an exposure on pulsar, a satellite can only record a series of events, which include the photon events from pulsar and the background noise events from X-ray detectors and the universe \cite{SONG2022203}. When there are sufficient photon events, we can estimate a pulse phase by handling the events, and estimate the position and velocity of the satellite via pulse phases \cite{Emadzadeh_2010}. However, the estimation of pulse phase is complicated because that there is no way to distinguish which event is a photon and that the count rate of photon event is usually much less than the count rate of background noise event. If the exposure on a pulsar is too short, the photons will be submerged by the background noise. Moreover, the count rate of background noise event varies with the type of X-ray telescope. There are currently two types of X-ray telescope, including the X-ray focusing telescope employed by \textit{NICER} and the X-ray collimated telescope employed by \textit{Insight}-HXMT. Given that the count rate of background noise event for X-ray collimated telescope is higher than that for X-ray focusing telescope \cite{Zhang2017}, \textit{Insight}-HXMT has to accumulate events much more than \textit{NICER} in order to have a pulse phase, which is as accurate as the pulse phase estimated with the data from \textit{NICER}. On the other hand, satellites perform the orbit motion through the whole exposure, which causes the frequency of pulsar signal vary with time. It makes the pulse phase estimation problem more difficult. To address this problem, \cite{Winternitz_2015} and \cite{Wang_2016} assume an approximation to the pulse phase evolution that captures most of the orbit dynamics and then correct this approximation by fitting a linear polynomial. This fit is accomplished by a two-dimensional grid search. When there are $N_{ph}$ events, the computational complexity of the grid search is about $O(N_{ph}N_{f}N_{q})$ where  $N_{f} \times N_{q}$ is the size of the grid \cite{Emadzadeh_2011}. This approach has been successfully applied to the \textit{NICER} onboard demonstration \cite{Winternitz2016}. 

Crab pulsar is an appealing source for XNAV with a small detector system and can provide a pulse phase estimation result more accurate than millisecond pulsars when the Crab pulsar and millisecond pulsars are exposed for the same exposure. However, Crab pulsar has a long spinning period and locates within a nebula, which would cause additional background noise  \cite{Tuo2019}. For Crab pulsar data from \textit{NICER}, the count rate of the photon event is only about 660 counts/s, but the count rate of the background noise event is about 13860 counts/s \cite{Ray2017}. In contrast, the whole count rate of the millisecond pulsar PSR B1937+21 data from \textit{NICER}, which was employed in the \textit{NICER} onboard demonstration, is only about 0.269 counts/s \cite{Ray2017}. According to \cite{Ray2017}, when there is an exposure on PSR B1937+21 lasting for 1000 s, the pulse phase estimation result for PSR B1937+21 can be as accurate as the pulse phase estimation result for Crab pulsar with an exposure of 191 s. Even in this case, the computational burden of the two-dimensional grid search for Crab pulsar is about 10364 times higher than that for PSR B1937+21. In fact, \textit{NICER} did not accomplish an onboard XNAV demonstration using Crab pulsar, but complemented the experiment on the ground \cite{Mitchell_2018, Winternitz2018}. Therefore, a computationally efficient pulse phase estimation method for Crab pulsar is needed. 

To this end, this paper proposes a fast on-orbit pulse phase estimation of Crab pulsar called X-ray pulsar navigaTion usIng on-orbiT pulsAr timiNg (XTITAN). We first derive pulse phase propagation models for real Crab pulsar data from \textit{Insight}-HXMT and \textit{NICER} respectively, given that \textit{Insight}-HXMT has to accumulate more events than \textit{NICER}. Then, one exposure is divided into several sub-exposures, and the pulse phases at the initial times of each segment are estimated via the prior knowledge of pulse phase propagation model. Moreover, those pulse phases are employed to fit the pulse phase propagation model again. When the iteration converges, the final pulse phase at the initial time of the whole exposure is employed for navigation. The pulse phase propagation model can be viewed as an on-orbit timing model for pulsar signal, and thus the method is called on-orbit pulsar timing. Compared with the pulse phase estimation method employed by \textit{NICER}, which is described in \cite{Winternitz_2015}, XTITAN also first approximates the pulse phase evolution with the aid of orbit dynamics of satellite, but corrects the approximation by performing an on-orbit pulsar timing instead of the grid search. As will be illustrated in Section \ref{sect.cca}, the computational complexity of XTITAN is much less than the two-dimensional grid search. In addition, an improved XTITAN is proposed, which iteratively estimates the position and velocity of satellite at the initial time of the exposure and performs on-orbit pulsar timing. As will be shown in the remainder of paper, for the \textit{NICER} data, XTITAN is about 58 times faster than the two-dimensional grid search, and thus is more suitable for the future onboard computation for Crab pulsar. In addition, when there are many exposures available, sequential employment of XTITAN at every exposure can provide a sequential navigation result.

The organization of the paper proceeds as follows. Section \ref{sect.PPM} derives the pulse phase propagation model for real Crab pulsar data. Section \ref{sect.OPT} shows the on-orbit pulsar timing method, and discusses its computational complexity. Section \ref{sect.Improved} improves the on-orbit pulsar timing by iteratively estimating the initial position and velocity of satellite at each exposure. Section \ref{Sect.ER} verifies the proposed algorithm by employing the real Crab data obtained from \textit{Insight}-HXMT and \textit{NICER}. 

\section{Pulse Phase Propagation Model Considering Satellite Orbital Motion}
\label{sect.PPM}
Assume the whole navigation process contains $\bar{N}$ exposures on the Crab pulsar, and the $j$th exposure starts at $t_{0}^{j}$ and ends at $t_{N}^{j}$. The events collected in the exposure is denoted as $\{t_{i}^{j}\}_{i=1}^{N}$. In order to estimate the pulse phase, every element of $\{t_{i}\}_{i=1}^{N}$ has to be corrected to the solar system barycenter (SSB) by \cite{Edwards_2006}
\begin{equation}
  \label{BC}
  \begin{aligned}
      t_{\mathrm{SSB}, i}^{j}&=g(t_{i}^{j})=t_{i}^{j}+\frac{1}{c}\boldsymbol{n}\bullet\left(\boldsymbol{r}(t_{i}^{j})+\boldsymbol{r}_{E}(t_{i}^{j})\right)\\
      &+2 \sum_{k} \frac{G M_{k}}{c^{3}} \ln \left(\boldsymbol{n} \bullet \boldsymbol{r}_{k}(t_{i}^{j})+\left\|\boldsymbol{r}_{k}(t_{i}^{j})\right\|\right)+\mathrm{H.O.T}
  \end{aligned}
,
\end{equation}
where $\boldsymbol{r}(t_{i}^{j})$ denotes the position of satellite relative to the Earth, $\boldsymbol{r}_{E}(t_{i}^{j})$ denotes the position of Earth with respective to the SSB, $\boldsymbol{n}$ denotes the direction vector of the pulsar, $ M_{k}$ is the mass of the $k$th celestial body and $\boldsymbol{r}_{k}(t_{i}^{j})$ is its position relative to the satellite, $c$ is the speed of light, and the $\mathrm{H.O.T}$ indicates the high-order term that can be ignored.

Assuming the pulse phase at $t_{\mathrm{SSB}, i}^{j}$ is $\phi_{0}$, the pulse phase at $t_{i}^{j}$, $\phi(t_{i}^{j})$, can be expressed as
\begin{equation}
\label{phapropag}
    \begin{aligned}
    \phi(t_{i}^{j})&=\phi_{0}+\nu_{0}\left[g(t_{i}^{j})-g(t_{0}^{j})\right]+\frac{1}{2} \dot{\nu}_{0}\left[g(t_{i}^{j})-g(t_{0}^{j})\right]^{2},
    \end{aligned}
\end{equation}
where $\phi_{0}$, $\nu_{0}$ and $\dot{\nu}_{0}$ are the phase, frequency of pulsar signal and its time derivative at $t_{\mathrm{SSB}, i}^{j}$, respectively.

Then, the frequency at $t_{i}^{j}$, $\nu(t_{i}^{j})$, can be derived as
\begin{equation}
    \label{eq.freqpropag}
    \begin{aligned}
    \nu(t_{i}^{j})=\frac{\mathrm{d}\phi(t_{i}^{j})}{\mathrm{d} t_{i}^{j}}=\left[
      \begin{aligned}
          \nu_{0}+\dot{\nu}_{0}\left(g(t_{i}^{j})-g(t_{0}^{j})\right)
      \end{aligned}
      \right]\frac{\mathrm{d}g(t_{i}^{j})}{\mathrm{d} t_{i}^{j}},
    \end{aligned}
\end{equation}
where
\begin{equation}
\begin{aligned}
  \frac{\mathrm{d}g(t_{i}^{j})}{\mathrm{d} t_{i}^{j}}&=1+\frac{1}{c}\boldsymbol{n}\bullet\left(\boldsymbol{v}(t_{i}^{j})+\boldsymbol{v}_{E}(t_{i}^{j})\right)\\
  &+2\sum_{k}\frac{G M_{k}}{c^{3}}\frac{\left[\left\|\boldsymbol{r}_{k}(t_{i}^{j})\right\|\boldsymbol{n}+\boldsymbol{r}_{k}(t_{i}^{j})\right]\bullet\boldsymbol{v}_{k}(t_{i}^{j})}{\left|\boldsymbol{n} \bullet \boldsymbol{r}_{k}(t_{i}^{j})+\left\|\boldsymbol{r}_{k}(t_{i}^{j})\right\|\right|\left\|\boldsymbol{r}_{k}(t_{i}^{j})\right\|}  
\end{aligned},
\end{equation}
where $\boldsymbol{v}(t_{i}^{j})$ denotes the velocity of satellite relative to the Earth, $\boldsymbol{v}_{E}(t_{i}^{j})$ denotes the velocity of Earth with respective to the SSB and $\boldsymbol{v}_{k}(t_{i}^{j})$ denotes the velocity of satellite with respective to the $k$th celestial body.

As illustrated in (\ref{phapropag}) and (\ref{eq.freqpropag}), the pulse phase evolution at an orbiting satellite is modulated by $\boldsymbol{r}(t_{i}^{j})$ and $\boldsymbol{v}(t_{i}^{j})$. However, in an autonomous navigation task, $\boldsymbol{r}(t_{i}^{j})$ and $\boldsymbol{v}(t_{i}^{j})$ are unknown. 

In order to estimate the pulse phase, we introduce the orbit dynamics of satellite into the pulse phase propagation model. Most time, the rough knowledge on $\boldsymbol{r}$ and $\boldsymbol{v}$ at $t_{0}^{j}$, $\tilde{\boldsymbol{r}}(t_{0}^{j})$ and $\tilde{\boldsymbol{v}}(t_{0}^{j})$, can be available by various means such as propagating the orbit dynamics model of satellite from the final epoch of the last exposure to $t_{0}^{j}$. In this case, the predicted positions and velocities of satellite at $\left\{t_{i}^{j}\right\}_{i=1}^{N}$, denoted as $\left\{\tilde{\boldsymbol{r}}(t_{i}^{j})\right\}_{i=1}^{N}$ and $\left\{\tilde{\boldsymbol{v}}(t_{i}^{j})\right\}_{i=1}^{N}$, can be obtained by propagating the orbit dynamics model which is initialized with $\tilde{\boldsymbol{r}}(t_{0}^{j})$ and $\tilde{\boldsymbol{v}}(t_{0}^{j})$.

Thus, we can linearize (\ref{phapropag}) around $\tilde{\boldsymbol{r}}(t_{i}^{j})$ and $\tilde{\boldsymbol{r}}(t_{0}^{j})$, leading to
\begin{equation}
  \label{phapropag_linear}
      \begin{aligned}
      \phi(t_{i}^{j})&=\phi_{0}+\tilde{\phi}(t_{i}^{j})+\left[\nu_{0}+\dot{\nu}_{0}\left(g(\tilde{t}_{i}^{j})-g(\tilde{t}_{0}^{j})\right)\right]\\
      &\bullet\left(\boldsymbol{G}_{i}\delta\boldsymbol{r}(t_{i}^{j})-\boldsymbol{G}_{0}\delta\boldsymbol{r}(t_{0}^{j})\right),
      \end{aligned}
\end{equation}
where
\begin{subequations}
  \label{subgtilde}
  \begin{align}
    \tilde{\phi}(t_{i}^{j})&=\nu_{0}\left[g(\tilde{t}_{i}^{j})-g(\tilde{t}_{0}^{j})\right]+\frac{1}{2} \dot{\nu}_{0}\left[g(\tilde{t}_{i}^{j})-g(\tilde{t}_{0}^{j})\right]^{2}\\
    \boldsymbol{G}_{i}&=\left.\frac{\partial g(t_{i}^{j})}{\partial  \boldsymbol{r}(t_{i}^{j})}\right|_{\boldsymbol{r}(t_{i}^{j})=\tilde{\boldsymbol{r}}(t_{i}^{j})}\\
    &=\frac{1}{c}\boldsymbol{n}+2\sum_{k}\frac{G M_{k}}{c^{3}}\frac{\boldsymbol{n}\left\|\tilde{\boldsymbol{r}}_{k}(t_{i}^{j})\right\|+\tilde{\boldsymbol{r}}_{k}(t_{i}^{j})}{\left|\boldsymbol{n} \bullet \tilde{\boldsymbol{r}}_{k}(t_{i}^{j})+\left\|\tilde{\boldsymbol{r}}_{k}(t_{i}^{j})\right\|\right|\left\|\tilde{\boldsymbol{r}}_{k}(t_{i}^{j})\right\|}\nonumber
  \end{align}
\end{subequations}
and $\delta\boldsymbol{r}(t_{i}^{j})$ is the error within $\tilde{\boldsymbol{r}}(t_{i}^{j})$.

As shown in \cite{Wang_2016}, $\delta\boldsymbol{r}(t_{i}^{j})$ can be expressed as a linear function of $\delta\boldsymbol{r}(t_{0}^{j})$ and $\delta\boldsymbol{v}(t_{0}^{j})$, i.e.,
\begin{equation}
  \label{eq.rssb}
    \delta\boldsymbol{r}(t_{i}^{j})=\boldsymbol{\Phi}_{rr}(t_{i}^{j}, t_{0}^{j})\delta\boldsymbol{r}(t_{0}^{j})+\boldsymbol{\Phi}_{rv}(t_{i}^{j}, t_{0}^{j})\delta\boldsymbol{v}(t_{0}^{j}).
\end{equation}

Substituting (\ref{eq.rssb}) into (\ref{phapropag_linear}) yields
\begin{equation}
  \label{phapropag_linear1}
      \begin{aligned}
      \phi(t_{i}^{j})&=\phi_{0}+\tilde{\phi}(t_{i}^{j})+\left[\nu_{0}+\dot{\nu}_{0}\left(g(\tilde{t}_{i}^{j})-g(\tilde{t}_{0}^{j})\right)\right]\\
      &\bullet\left[
        \begin{aligned}
          \left(\boldsymbol{G}_{i}\boldsymbol{\Phi}_{rr}(t_{i}^{j}, t_{0}^{j})-\boldsymbol{G}_{0}\right)\delta\boldsymbol{r}(t_{0}^{j})\\+\boldsymbol{G}_{i}\boldsymbol{\Phi}_{rv}(t_{i}^{j}, t_{0}^{j})\delta\boldsymbol{v}(t_{0}^{j})
        \end{aligned}
          \right].
      \end{aligned}
\end{equation}

$\boldsymbol{\Phi}_{rr}(t_{i}^{j}, t_{0}^{j})$ and $\boldsymbol{\Phi}_{rv}(t_{i}^{j}, t_{0}^{j})$ in (\ref{phapropag_linear1}) can both be expanded as a polynomial of $t_{i}^{j}-t_{0}^{j}$, i.e., 
\begin{subequations}
\label{eq.transt}
\begin{align}
  \boldsymbol{\Phi}_{rr}(t_{i}^{j}, t_{0}^{j})&=\boldsymbol{I}_{3 \times 3}+\sum_{m=1}^{\infty} \frac{1}{m !} \boldsymbol{\varphi}_{m}\left(t_{i}^{j}-t_{0}^{j}\right)^{m}\\
  \boldsymbol{\Phi}_{rv}(t_{i}^{j}, t_{0}^{j})&=\sum_{m=1}^{\infty} \frac{1}{m!} \boldsymbol{\gamma}_{m}\left(t_{i}^{j}-t_{0}^{j}\right)^{m}
\end{align}
\end{subequations}
where $\boldsymbol{\varphi}_{m}$ and $\boldsymbol{\gamma}_{m}$ are constant matrices. 

Substituting (\ref{eq.transt}) into (\ref{phapropag_linear1}), (\ref{phapropag_linear1}) becomes
\begin{equation}
  \label{eq.phapropag_linear2}
  \begin{aligned}
    \phi(t_{i}^{j})&=\phi_{0}+\tilde{\phi}(t_{i}^{j})+\left[\nu_{0}+\dot{\nu}_{0}\left(g(\tilde{t}_{i}^{j})-g(\tilde{t}_{0}^{j})\right)\right]\\
    &\bullet\left[
      \begin{aligned} 
        \left(\boldsymbol{G}_{i}+\sum_{m=1}^{\infty} \frac{1}{m !} \boldsymbol{G}_{i}\boldsymbol{\varphi}_{m}\left(t_{i}^{j}-t_{0}^{j}\right)^{m}-\boldsymbol{G}_{0}\right)\delta\boldsymbol{r}(t_{0}^{j})\\
         +\sum_{m=1}^{\infty} \frac{1}{m!} \boldsymbol{G}_{i}\boldsymbol{\gamma}_{m}\delta\boldsymbol{v}(t_{0}^{j})\left(t_{i}^{j}-t_{0}^{j}\right)^{m}
      \end{aligned}
    \right].
    \end{aligned}  
\end{equation}

In order to simplify (\ref{eq.phapropag_linear2}), we exploit the relationship between $\boldsymbol{G}_{i}$ and $\boldsymbol{G}_{0}$. In (\ref{subgtilde}b), 
\begin{equation}
  \tilde{\boldsymbol{r}}_{k}(t_{i}^{j})=\tilde{\boldsymbol{r}}_{\mathrm{SC/E}, k}(t_{i}^{j})+\tilde{\boldsymbol{r}}_{\mathrm{E}, k}(t_{i}^{j})-\tilde{\boldsymbol{p}}_{k}(t_{i}^{j}),  
\end{equation}
where $\tilde{\boldsymbol{r}}_{\mathrm{SC/E}, k}(t_{i}^{j})$ is the predicted position of the satellite relative to the Earth at $t_{i}^{j}$, $\tilde{\boldsymbol{r}}_{\mathrm{E}, k}(t_{i}^{j})$ is the position of the Earth relative to the SSB at $t_{i}^{j}$ and $\tilde{\boldsymbol{p}}_{k}(t_{i}^{j})$ denotes the position of the $k$th celestial body relative to the SSB at $t_{i}^{j}$. 

Although it seems the second term on the right side of (\ref{subgtilde}b) should consider the impact of all the celestial bodies in the solar system, only the Sun and the Jupiter are considered in real applications because the sum of their mass accounts for about 99\% of the whole mass of the solar system. Given that the distance between the Sun and the Earth is about $1.496\times 10^{8}$ km and that the distances between the satellites, which include the ISS and the \textit{Insight}-HXMT, and the Earth is about 500 km, we have $\tilde{\boldsymbol{r}}_{k}(t_{i}^{j}) \approx\tilde{\boldsymbol{r}}_{\mathrm{E}, k}(t_{i}^{j})-\tilde{\boldsymbol{p}}_{k}(t_{i}^{j})$. An exposure typically lasts for several hundred to 3000 s, during which the Sun, the Earth and the Jupiter can be approximated to be stationary. Thus, $\tilde{\boldsymbol{r}}_{k}(t_{i}^{j}) \approx \tilde{\boldsymbol{r}}_{k}(t_{0}^{j})$, and $\boldsymbol{G}_{i}\approx \boldsymbol{G}_{0}$. In this case, (\ref{eq.phapropag_linear2}) becomes
\begin{equation}
  \label{eq.phapropag_linear3}
  \phi(t_{i}^{j})=\phi_{0}+\tilde{\phi}(t_{i}^{j})+
  \sum_{m=1}^{\infty}\bar{\nu}_{m}\left(t_{i}^{j}-t_{0}^{j}\right)^{m},
\end{equation} 
where
\begin{equation}
  \label{eq.phapropag_linear4}
  \begin{aligned}
      \bar{\nu}_{m}&= \frac{1}{m !}\left[\nu_{0}+\dot{\nu}_{0}\left(g(\tilde{t}_{i}^{j})-g(\tilde{t}_{0}^{j})\right)\right]\\
      &\bullet\left[\boldsymbol{G}_{i}\boldsymbol{\varphi}_{m}\delta\boldsymbol{r}(t_{0}^{j})
    +\boldsymbol{G}_{i}\boldsymbol{\gamma}_{m}\delta\boldsymbol{v}(t_{0}^{j})\right].
  \end{aligned}
\end{equation} 

The value of $m$ depends on the duration of the $j$th exposure and on the orbit altitude of satellite. As will be shown in the section \ref{sect.OPT}, in order to fulfill XTITAN, one exposure has to be divided into several sub-exposures, the duration of which should ensure one pulse phase can be estimated. It is because that the pulse phase estimation would fail if the exposure is too short to collect sufficient photon events. We found that an effective exposure for \textit{Insight}-HXMT and for \textit{NICER} should be at least 2000 s and 1000 s, respectively. In this case, $m$ for the data from \textit{Insight}-HXMT should be 2, and $m$ for the data from \textit{NICER} should be 1. Finally, the phase propagation models for \textit{Insight}-HXMT and for \textit{NICER} are
\begin{subequations}
  \label{phapropag_ploy}
      \begin{align}
      \phi_{\mathrm{HXMT}}(t_{i}^{j})&=\tilde{\phi}(t_{i}^{j})+\phi_{0}+\bar{\nu}_{1}(t_{i}^{j}-t_{0}^{j})+\bar{\nu}_{2}(t_{i}^{j}-t_{0}^{j})^{2}\\
      \phi_{\mathrm{NICER}}(t_{i}^{j})&=\tilde{\phi}(t_{i}^{j})+\phi_{0}+\bar{\nu}_{1}(t_{i}^{j}-t_{0}^{j}),
      \end{align}
\end{subequations}  
where $\bar{\nu}_{1}$ and $\bar{\nu}_{2}$ are hyperparameters that are needed to be estimated along with $\phi_{0}$.

There are only one or two hyperparameters in (\ref{phapropag_ploy}). In contrast, if (\ref{phapropag}) is employed to estimate the pulse phase, $\boldsymbol{r}(t_{i})$ and $\boldsymbol{v}(t_{i})$ have to be approximated by a piece-wise linear model which involves numerous hyperparameters\cite{Wang_2016}. 

In \cite{Wang_2016}, we derived a pulse phase propagation model similar to (\ref{phapropag_ploy}). However, the derivation in this paper is more rigorous than \cite{Wang_2016}. There are two reasons: 1) \cite{Wang_2016} only considers the Romer delay in the barycenter correction, in contrast, (\ref{BC}) considers the Romer delay and the Shapiro delay; and 2) the phase evolution model in \cite{Wang_2016} only considers the frequency of pulsar signal, in contrast, (\ref{phapropag_ploy}) is derived from (\ref{phapropag}), which contains not only the frequency of pulsar signal but the time derivative of frequency.  

\section{On-orbit Pulsar Timing for Estimating $\phi_{0}$, $\bar{\nu}_{1}$ and $\bar{\nu}_{2}$}
\label{sect.OPT}
\subsection{Motivation}
\label{subsect.mot}
To estimate $\phi_{0}$ in (\ref{phapropag_ploy}), the most famous method is the maximum likelihood estimator (MLE). Based on that the events follow an inhomogeneous Poisson process and (\ref{phapropag_ploy}), a log-likelihood function of $\{\phi(t_{i}^{j})\}_{i=1}^{N}$ can be expressed as \cite{Emadzadeh_2011}
\begin{subequations}
\label{eq.dphase1}
\begin{align}
 LLF_{\mathrm{HXMT}}=\sum_{i=1}^{N} \ln \left(\lambda\left(\phi_{\mathrm{HXMT}}\left(t_{i}^{j};\phi_{0}, \bar{\nu}_{1}, \bar{\nu}_{2}\right)\right)\right)\\
 LLF_{\mathrm{NICER}}=\sum_{i=1}^{N} \ln \left(\lambda\left(\phi_{\mathrm{NICER}}\left(t_{i}^{j};\phi_{0}, \bar{\nu}_{1}\right)\right)\right)
\end{align}
\end{subequations} 
where
\begin{equation}
\label{eq.dphase2}
\lambda(t)=\alpha h\left(\phi(t)\right)+\beta
\end{equation}
with $h\left(\bullet\right)$  the pulsar profile template, $\alpha$ and $\beta$  the detected rate constants.

$\phi_{0}$, $\bar{\nu}_{1}$ and $\bar{\nu}_{2}$ in (\ref{eq.dphase1}) are estimated by solving the minimization problem of
\begin{subequations}
  \label{eq.LLF}
  \begin{align}
      \hat{\phi}_{0}, \hat{\bar{\nu}}_{1}, \hat{\bar{\nu}}_{2}= \arg \min_{\phi_{0}, \bar{\nu}_{1}, \bar{\nu}_{2}}LLF_{\mathrm{HXMT}}\\
      \hat{\phi}_{0}, \hat{\bar{\nu}}_{1}= \arg \min_{\phi_{0}, \bar{\nu}_{1}}LLF_{\mathrm{NICER}}.
  \end{align}
\end{subequations}

\textit{NICER} employed the two-dimensional grid search to solve (\ref{eq.LLF}b). For clarity, the procedure of two-dimensional grid search is shown as \textbf{Algorithm for Comparison 1}. It indicates the computational complexity of the two-dimensional grid search is about $O(NN_{\phi_{0}}N_{\bar{\nu}_{1}})$ ($N_{\phi_{0}}$ and $N_{\bar{\nu}_{1}}$ are the number of grid nodes) \cite{wang_zheng_2016}. As mentioned in Section \ref{sect:intro}, if the exposure on Crab pulsar lasts for 1000 s, $N$ would be $1.442\times 10^{7}$. When $N_{\phi_{0}}$ and $N_{\bar{\nu}_{1}}$ are both set as 1000, the computational complexity is about $O(1.442\times 10^{10})$.

\begin{table}[h]\label{algorithm3}
  \begin{tabular}{l}
  \hline
      \textbf{Algorithm for Comparison 1:} \\
      ~~~~~~~~~~Two-dimensional Grid Search for $\phi_{0}$ and $\bar{\nu}_{1}$\\
       \hline
      1: \textbf{Initialization}: \\
      2: ~~~~Assume the search spaces for $\phi_{0}$ and $\bar{\nu}_{1}$ are [0, 1)\\ 
      ~~~~~~~~and $[\bar{\nu}_{1,\mathrm{min}}, \bar{\nu}_{1,\mathrm{max}}]$ respectively.\\
      3: ~~~~Divide [0, 1) into $N_{\phi_{0}}$ segments,\\
      ~~~~~~~and divide $[\bar{\nu}_{1,\mathrm{min}}, \bar{\nu}_{1,\mathrm{max}}]$ into $N_{\nu_{1}}$ segments.\\
      4: ~~~~Design a $N_{\phi_{0}} \times N_{\nu_{1}}$ grid;\\
      5: \textbf{for} $k=1,\cdots, N_{\phi_{0}}$ \textbf{do}\\
      6: ~~~~$\phi_{0}^{(k)}=\frac{k-1}{N_{\phi_{0}}}$\\
      7: ~~~~\textbf{for} $l=1,\cdots, N_{\nu_{1}}$ \textbf{do}\\
      8: ~~~~~~~~$\bar{\nu}_{1}^{(l)}=\bar{\nu}_{1,\mathrm{min}}+\frac{l-1}{N_{\nu_{1}}}\left(\bar{\nu}_{1,\mathrm{max}}-\bar{\nu}_{1,\mathrm{min}}\right)$\\
      9: ~~~~~~~~\textbf{for} $i=1,\cdots, N$ \textbf{do}\\
      10: ~~~~~~~~~~~~~~Calculate $LLF(k, l, i)=\ln \left(\lambda\left(\phi\left(t_{i}^{j};\phi_{0}^{(k)}, \bar{\nu}_{1}^{(l)}\right)\right)\right)$\\
      11: ~~~~~~~~\textbf{end for}\\
      12: ~~~~\textbf{end for}\\
      13:\textbf{end for}\\
      14: $\hat{\phi}_{0}, \hat{\bar{\nu}}_{1}= \arg \min LLF$\\
      15:\textbf{Output:} $\hat{\phi}_{0}, \hat{\bar{\nu}}_{1}$\\ 
      \hline
  \end{tabular}
\end{table}

\subsection{Framework}
\label{sect.framwork}
In order to reduce the computation complexity of pulse phase estimation, we circumvent the MLE, and propose the on-orbit pulsar timing method to iteratively estimate $\phi_{0}$, $\bar{\nu}_{1}$ and $\bar{\nu}_{2}$. For simplicity, in the remainder of this paper, we derive XTITAN based on (\ref{phapropag_ploy}a). The investigation is also feasible when (\ref{phapropag_ploy}b) is employed.

It can be learned from (\ref{phapropag_ploy}a), $\phi(t_{i}^{j})$ is a function of $\tilde{\boldsymbol{r}}(t_{i}^{j})$, $\tilde{\boldsymbol{v}}(t_{i}^{j})$, $\phi_{0}$, $\bar{\nu}_{1}$ and $\bar{\nu}_{2}$. Moreover, $\tilde{\boldsymbol{r}}(t_{i}^{j})$ and $\tilde{\boldsymbol{v}}(t_{i}^{j})$ can be derived from propagating $\boldsymbol{r}(t_{0}^{j})$ and $\boldsymbol{v}(t_{0}^{j})$. When $\tilde{\boldsymbol{r}}(t_{0}^{j})$ and $\tilde{\boldsymbol{v}}(t_{0}^{j})$ are given, $\phi(t_{i}^{j})$ depends on $\phi_{0}, \bar{\nu}_{1}, \bar{\nu}_{2}$ which are constant through the $j$th exposure. It indicates that (\ref{phapropag_ploy}) not only can be viewed as a phase propagation model but also a timing model. Thus, we can estimate $\phi_{0}, \bar{\nu}_{1}, \bar{\nu}_{2}$ by fitting the timing model. That is the very reason for the name of the proposed method. 

If the whole $j$th exposure is divided into $M$ sub-exposures and the start time at the $l$th sub-exposure is $\tau_{l}$, we have
\begin{equation}
  \label{eq.PulsarTiming1}
  \boldsymbol{\phi}=\tilde{\boldsymbol{\phi}}+\boldsymbol{\theta}\bar{\boldsymbol{\nu}},
\end{equation}
where
\begin{subequations}
  \begin{align}
    \bar{\boldsymbol{\nu}}&=\left[\phi_{0}, \bar{\nu}_{1}, \bar{\nu}_{2}\right]^{\mathrm{T}}\\
    \boldsymbol{\phi}&=\left[\phi\left({\tau_{1}}\right), \phi\left({\tau_{2}}\right), \cdots, \phi\left({\tau_{M}}\right)\right]^{\mathrm{T}}\\
    \tilde{\boldsymbol{\phi}}&=\left[\tilde{\phi}\left({\tau_{1}}\right), \tilde{\phi}\left({\tau_{2}}\right), \cdots, \tilde{\phi}\left({\tau_{M}}\right)\right]^{\mathrm{T}}\\
    \boldsymbol{\theta} &=\left[\begin{array}{ccc }
     1 & \tau_{1}-t_{0}^{j} & \left(\tau_{1}-t_{0}^{j}\right)^{2}\\
     1 & \tau_{2}-t_{0}^{j} & \left(\tau_{2}-t_{0}^{j}\right)^{2}\\
     \vdots & \vdots & \vdots\\
     1 & \tau_{M}-t_{0}^{j} & \left(\tau_{M}-t_{0}^{j}\right)^{2}
      \end{array}\right]
  \end{align}
\end{subequations}

The estimate of $\bar{\boldsymbol{\nu}}$ can be obtained by solving the following optimization problem,
\begin{equation}
  \label{eq.cost_func_ls}
  \hat{\bar{\boldsymbol{\nu}}}= \arg \min_{\bar{\boldsymbol{\nu}}}\left\|\boldsymbol{\phi}-\tilde{\boldsymbol{\phi}}-\boldsymbol{\theta}\bar{\boldsymbol{\nu}}\right\|.
\end{equation}

Equation (\ref{eq.cost_func_ls}) is commonly solved by the standard least square algorithm, leading to
\begin{subequations}
  \label{eq.solu_1}
  \begin{align}
    \hat{\bar{\boldsymbol{\nu}}}&=\boldsymbol{\vartheta}\left(\boldsymbol{\phi}-\tilde{\boldsymbol{\phi}}\right)\\
    \boldsymbol{\vartheta} &= \left(\boldsymbol{\theta}^{\mathrm{T}} \boldsymbol{\theta}\right)^{-1} \boldsymbol{\theta}^{\mathrm{T}}.
  \end{align}
\end{subequations}

As shown in (\ref{eq.solu_1}), $\boldsymbol{\vartheta}$ is constant when 
$\left\{\tau_{i}\right\}_{i=1}^{M}$ are given. However, if the matrix $\boldsymbol{\theta}^{\mathrm{T}} \boldsymbol{\theta}$ is approximately ill-conditioned, we cannot have a reliable inverse of $\boldsymbol{\theta}^{\mathrm{T}} \boldsymbol{\theta}$ and thus $\boldsymbol{\vartheta}$ is inaccurate. In this case, we can exploit the prior information on $\bar{\boldsymbol{\nu}}$, and modify the cost function in (\ref{eq.cost_func_ls}) to be a regularized one,
\begin{equation}
  \label{eq.cost_func_ls_reg}
  \hat{\bar{\boldsymbol{\nu}}}= \arg \min_{\bar{\boldsymbol{\nu}}}\left(\left\|\boldsymbol{\phi}-\tilde{\boldsymbol{\phi}}-\boldsymbol{\theta}\bar{\boldsymbol{\nu}}\right\|+\gamma\|\bar{\boldsymbol{\nu}}\|\right),
\end{equation}
\noindent
where $\gamma$ is the hyperparameter that is needed to be determined.

  The solution of (\ref{eq.cost_func_ls_reg}) is
\begin{subequations}
  \label{eq.solu_2}
  \begin{align}
    \hat{\bar{\boldsymbol{\nu}}}&=\bar{\boldsymbol{\vartheta}}\left(\boldsymbol{\phi}-\tilde{\boldsymbol{\phi}}\right)\\
    \bar{\boldsymbol{\vartheta}} &= \left(\boldsymbol{\theta}^{\mathrm{T}} \boldsymbol{\theta}+\gamma\boldsymbol{I}\right)^{-1} \boldsymbol{\theta}^{\mathrm{T}}.
  \end{align}
\end{subequations}
\noindent
where $\boldsymbol{I}$ denotes the unit matrix. When the $\gamma$ is properly selected, $\boldsymbol{\theta}^{\mathrm{T}} \boldsymbol{\theta}+\gamma\boldsymbol{I}$ is always invertible.  

To further save the computational burden, in the $l$th ($l=1,2,\cdots, M$) sub-exposure, we apply the general epoch folding (GEF) to recover an empirical profile and to estimate $\phi_{l}$ by comparing the empirical profile with the template.

\subsubsection{General Epoch Folding}
\label{GEF}
The epoch folding has been widely employed to recover the empirical profile of pulsar. The classical epoch folding directly employs the event series to recover an empirical profile, which is defined within [0, $P$) with $P$ of the pulsar signal's period \cite{Emadzadeh_2011}. However, as shown in (\ref{phapropag_ploy}), the frequency of pulsar signal is time-varying in real applications and so does the period of pulsar's signal. In this case, if the empirical profile is still defined in the [0, $P$), the resulting empirical profile will be smeared. Thus, we propose the general epoch folding (GEF) method.  

Take the event series $\left\{t_{i}\right\}_{i=1}^{N}$ for example. The procedure of GEF proceeds as follows. 1) GEF first applies (\ref{phapropag_ploy}) to each element of $\left\{t_{i}\right\}_{i=1}^{N}$ to obtain the phase series $\left\{\phi(t_{i})\right\}_{i=1}^{N}$, and equally divides the first cycle into $N_{b}$ bins. 2) The events, phases of which are more than one cycle, are folded back into the first one. 3) An empirical profile can be recovered by counting the photons dropping into each bin and by normalizing the number of photons.

Finally, the empirical profile in the $i$th bin ($i \in [1, N_{b}]$), $\rho_{\mathrm{E}}(\phi(i))$, can be described by
\begin{equation}
\label{eq.fsa3}
\rho_{\mathrm{E}}\left(\phi(i)\right)=\frac{n_{i}}{N_{p}},
\end{equation}
where $n_{i}$ is the number of events in the $i$th bin and $N_{p}$ is the number of all recorded events.

Compared with the classical epoch folding shown in \cite{Emadzadeh2011}, GEF can successfully recover the empirical profile even there is a quadratic term $\bar{\nu}_{2}(t_{i}-t_{0})^{2}$ in (\ref{phapropag_ploy}) because GEF employs $\left\{\phi(t_{i})\right\}_{i=1}^{N}$ instead of $\left\{t_{i}\right\}_{i=1}^{N}$. Moreover, GEF uses $N_{p}$ to normalize the empirical profile. In this way, the size of bin is constant, and thus the empirical profile is stable. In contrast, the classical epoch folding uses $T_{b}N_{fc}$, where $T_{b}=P/N_{b}$ and $N_{fc}$ is the number of pulsar period in the exposure, for normalization \cite{Emadzadeh_2011}. However, $T_{b}$ varies because that $P$ varies. Then, the size of bin varies and will cause the empirical profile smear.

\subsubsection{Brief Introduction of Pulse Phase Estimation}
We now briefly introduce the estimation of $\phi_{l}$ by comparing the empirical profile and the template. For someone who is interested, please find the detailed descriptions in \cite{Emadzadeh2011}. Assuming an empirical profile, $\boldsymbol{\rho}_{\mathrm{E}}$ can be represented as $\boldsymbol{\rho}_{\mathrm{E}}=\left[\rho_{\mathrm{E}}\left(\phi(1)\right), \rho_{\mathrm{E}}\left(\phi(2)\right), \cdots, \rho_{\mathrm{E}}\left(\phi(N_{b})\right)\right]^{\mathrm{T}}$. Meanwhile, the template can be also denoted as $\boldsymbol{\rho}_{\mathrm{T}}=\left[\rho_{\mathrm{T}}\left(\phi(1)\right), \rho_{\mathrm{T}}\left(\phi(2)\right), \cdots, \rho_{\mathrm{T}}\left(\phi(N_{b})\right)\right]^{\mathrm{T}}$. In this case, the estimate of $\phi_{l}$, $\hat{\phi}_{l}$, can be obtained by solving

\begin{equation}
  \label{eq.profile_compar}
    \hat{\phi}_{l} = \arg \min_{\phi_{l}} \left\|\boldsymbol{\rho}_{\mathrm{E}}-\boldsymbol{\rho}_{\mathrm{T}}\right\|.
\end{equation}

The classical methods to address (\ref{eq.profile_compar}) are cross-correlation \cite{Emadzadeh2011} and nonlinear least square (NLS) \cite{Emadzadeh_2010}. The Cramer-Rao Low Bounds (CRLBs) for the result of cross-correlation and NLS are derived in \cite{Emadzadeh_2010, Emadzadeh2011}.

\subsubsection{Summary of The Proposed Algorithm}
As illustrated in Section \ref{GEF}, it is needed to give an initial guess of $\bar{\boldsymbol{\nu}}$ for GEF and to estimate $\boldsymbol{\phi}$. The estimated $\boldsymbol{\phi}$ is employed to update $\bar{\boldsymbol{\nu}}$ again. Thus, $\hat{\bar{\boldsymbol{\nu}}}$ should be estimated in an iterated way.

When $\tilde{\boldsymbol{r}}(t_{0}^{j})$ and $\tilde{\boldsymbol{v}}(t_{0}^{j})$ are given for the $j$th exposure, the iterated procedure is summarized as \textbf{Algorithm 1}.
\begin{table}[h]\label{algorithm1}
  \begin{tabular}{l}
  \hline
      \textbf{Algorithm 1} Iterated Estimation of $\phi_{0}$, $\bar{\nu}_{1}$ and $\bar{\nu}_{2}$\\ \hline
      1: \textbf{Initialization}: \\
      ~~~~Divide the $j$th exposure into $M$ sub-exposures;\\
      ~~~~Set $\bar{\boldsymbol{\nu}}^{(0)}=\left[\phi_{0}^{(0)}, \bar{\nu}_{1}^{(0)}, \bar{\nu}_{2}^{(0)}\right]^{\mathrm{T}}$;\\
      2: \textbf{for} $k=1,\cdots, K$ \textbf{do}\\
      3: ~~~~\textbf{for} $l=1,\cdots, M$ \textbf{do}\\
      4: ~~~~~~~~Apply the GEF to $\left\{t_{i}\right\}_{i=1}^{N}$ to recover an empirical profile;\\
      5: ~~~~~~~~Estimate $\phi_{l}^{(k)}$ by comparing the empirical profile \\
      ~~~~~~~~~~~with the template;\\
      6: ~~~~\textbf{end for}\\
      7: ~~~~Estimate $\bar{\boldsymbol{\nu}}^{(k)}$ according to (\ref{eq.solu_1}) or (\ref{eq.solu_2});\\
      8: ~~~~\textbf{if} $\|\bar{\boldsymbol{\nu}}^{(k)}-\bar{\boldsymbol{\nu}}^{(k-1)}\|<\epsilon $\\
      9: ~~~~~~~~break;\\
      10:~~~~\textbf{else}\\
      11:~~~~~~~~$\bar{\boldsymbol{\nu}}^{(k+1)}\leftarrow\bar{\boldsymbol{\nu}}^{(k)}$;\\
      12:~~~~~~~~$k\leftarrow k+1$;\\
      13:~~~~\textbf{end if}\\
      14:\textbf{end for}\\
      15:\textbf{Output}: $\hat{\bar{\boldsymbol{\nu}}}=\left[\hat{\phi}_{0}, \hat{\bar{\nu}}_{1}, \hat{\bar{\nu}}_{2}\right]^{\mathrm{T}}$.\\ 
      \hline
  \end{tabular}
\end{table}

\subsection{Computational Complexity Analysis of \textbf{Algorithm 1}}
\label{sect.cca}
In one iteration, the computation burden of \textbf{Algorithm 1} is mainly spent on (\ref{eq.solu_1}), the computation complexity of which is about $O\left(27+9M\right)$, and on the profile comparison with the computation complexity about $O\left(N_{b}^{2}\right)$. Moreover, the matrix inverse in (\ref{eq.solu_1}) only needs to be performed once. It means the computational complexity of \textbf{Algorithm 1} is about $O\left(27+9M+KMN_{b}^{2}\right)$. For \textit{Insight}-HXMT and \textit{NICER} data, we found $M$ set as 6 and $K$ usually less than 3. When \textbf{Algorithm 1} is applied to the example provided in Section \ref{subsect.mot} and $N_{b}$ is set as 1000, the computational complexity is about $O(1.8\times 10^{6})$, which is about $10^{-4}$ of the computational complexity of two-dimensional grid search shown in Section \ref{subsect.mot}.

\section{Iterated On-orbit Pulsar Timing and Estimation of Satellite State}
\label{sect.Improved}

\textbf{Algorithm 1} iteratively estimates $\hat{\phi}_{0}$, $\hat{\bar{\nu}}_{1}$ and $\hat{\bar{\nu}}_{2}$ on the premise that $\tilde{\boldsymbol{r}}(t_{0}^{j})$ and $\tilde{\boldsymbol{v}}(t_{0}^{j})$ are given. The accuracies of $\tilde{\boldsymbol{r}}(t_{0}^{j})$ and $\tilde{\boldsymbol{v}}(t_{0}^{j})$ limit the estimation accuracies of $\hat{\phi}_{0}$, $\hat{\bar{\nu}}_{1}$ and $\hat{\bar{\nu}}_{2}$. Thus, we improve \textbf{Algorithm 1} to estimate $\phi_{0}$, $\bar{\nu}_{1}$, $\bar{\nu}_{2}$, $\boldsymbol{r}(t_{0}^{j})$ and $\boldsymbol{v}(t_{0}^{j})$ together.

From the viewpoint of pulsar timing, when $\boldsymbol{\phi}$ is obtained, we can have
\begin{equation}
  \label{eq.BOD}
  \left[\begin{array}{c}
    \phi_{1}\\
    \phi_{2}\\
    \vdots\\
    \phi_{M}
     \end{array}\right]=F_{0}\left[\begin{array}{c}
      g(\tau_{1})-T_{0}\\
      g(\tau_{2})-T_{0}\\
      \vdots\\
      g(\tau_{M})-T_{0}
       \end{array}\right]+\frac{F_{1}}{2}\left[\begin{array}{c}
        \left(g(\tau_{1})-T_{0}\right)^{2}\\
        \left(g(\tau_{2})-T_{0}\right)^{2}\\
        \vdots\\
        \left(g(\tau_{M})-T_{0}\right)^{2}
         \end{array}\right],
\end{equation}
where $F_{0}$ and $F_{1}$ are the spinning frequency of pulsar and its time derivative at $T_{0}$ respectively. $F_{0}$ and $F_{1}$ can be obtained from the public ephemeris of pulsar.

As shown in (\ref{BC}), $g(\tau_{l})$ ($l=1,2,\cdots, M$) is a function of $\boldsymbol{r}(\tau_{l})$. (\ref{eq.BOD}) can be rewritten as
\begin{equation}
  \label{eq.BOD1}
  \boldsymbol{\phi}=\left[\begin{array}{c}
    h_{1}(\boldsymbol{x}(\tau_{1}))\\
    h_{2}(\boldsymbol{x}(\tau_{2}))\\
    \vdots\\
    h_{M}(\boldsymbol{x}(\tau_{M}))
     \end{array}\right],
\end{equation}
where $\boldsymbol{x}=\left[\boldsymbol{r}^{\mathrm{T}}, \boldsymbol{v}^{\mathrm{T}}\right]^{\mathrm{T}}$, and
\begin{equation}
  h_{l}(\boldsymbol{x}(\tau_{l}))= F_{0}\left(g(\tau_{l})-T_{0}\right)+\frac{F_{1}}{2}\left((g(\tau_{l})-T_{0}\right)^{2}.
\end{equation}

When $\tilde{\boldsymbol{x}}(t_{0}^{j})=\left[\tilde{\boldsymbol{r}}(t_{0}^{j})^{\mathrm{T}}, \tilde{\boldsymbol{v}}(t_{0}^{j})^{\mathrm{T}}\right]^{\mathrm{T}}$ are given, the predicted states at $\tau_{l}$ ($l=1,2,\cdots, M$), $\tilde{\boldsymbol{x}}(\tau_{l})$, can be obtained by propagating the satellite orbit dynamics model initialized with $\tilde{\boldsymbol{x}}(t_{0}^{j})$. Equation (\ref{eq.BOD1}) can be linearized around $\tilde{\boldsymbol{x}}(\tau_{l})$, and becomes
\begin{equation}
  \label{eq.BOD2}
  \boldsymbol{\Delta\phi}=\left[\begin{array}{c}
    \boldsymbol{H}_{1}\delta\boldsymbol{x}(\tau_{1})\\
    \boldsymbol{H}_{2}\delta\boldsymbol{x}(\tau_{2})\\
    \vdots\\
    \boldsymbol{H}_{M}\delta\boldsymbol{x}(\tau_{M})
     \end{array}\right],
\end{equation}
where
\begin{subequations}
  \begin{align}
    \boldsymbol{\Delta\phi}&=\boldsymbol{\phi}-\left[\begin{array}{c}
      h_{1}(\tilde{\boldsymbol{x}}(\tau_{1}))\\
      h_{2}(\tilde{\boldsymbol{x}}(\tau_{2}))\\
      \vdots\\
      h_{M}(\tilde{\boldsymbol{x}}(\tau_{M}))
       \end{array}\right]\\
      \boldsymbol{H}_{l}&=\left.\frac{\partial h_{l}}{\partial \boldsymbol{x}}\right|_{\boldsymbol{x}=\tilde{\boldsymbol{x}}(\tau_{l})}\\
      \delta\boldsymbol{x}(\tau_{l})&=\boldsymbol{x}(\tau_{l})-\tilde{\boldsymbol{x}}(\tau_{l}).
  \end{align}
\end{subequations}

Meanwhile, $\delta\boldsymbol{x}(\tau_{l})$ can be expressed as
\begin{equation}
  \label{eq.Transit}
  \delta\boldsymbol{x}(\tau_{l})=\boldsymbol{\Phi}\left(\tau_{l}, t_{0}^{j}\right)\delta\boldsymbol{x}(t_{0}^{j}),
\end{equation} 
where $\boldsymbol{\Phi}\left(\tau_{l}, t_{0}^{j}\right)$ is the state transition matrix. $\boldsymbol{\Phi}\left(\tau_{l}, t_{0}^{j}\right)$ can be calculated by digital integral technique, which is introduced in detail in \cite{Tapley_2004}. 

Substituting (\ref{eq.Transit}) into (\ref{eq.BOD2}) yields
\begin{subequations}
  \label{eq.BOD3}
  \begin{align}
      \boldsymbol{\Delta\phi}&=\bar{\boldsymbol{H}}\delta\boldsymbol{x}(t_{0}^{j})\\
  \bar{\boldsymbol{H}}&=\left[\begin{array}{c}
    \boldsymbol{H}_{1}\boldsymbol{\Phi}\left(\tau_{1}, t_{0}^{j}\right)\\
    \boldsymbol{H}_{2}\boldsymbol{\Phi}\left(\tau_{2}, t_{0}^{j}\right)\\
    \vdots\\
    \boldsymbol{H}_{M}\boldsymbol{\Phi}\left(\tau_{M}, t_{0}^{j}\right)
     \end{array}\right]. 
  \end{align}
\end{subequations}

Thus,
\begin{equation}
  \label{eq.deltax}
  \hat{\delta\boldsymbol{x}}(t_{0}^{j})=\left(\bar{\boldsymbol{H}}^{\mathrm{T}}\bar{\boldsymbol{H}}\right)^{-1}\bar{\boldsymbol{H}}^{\mathrm{T}}\boldsymbol{\Delta\phi}.
\end{equation}

As illustrated in (\ref{eq.cost_func_ls})-(\ref{eq.solu_2}), (\ref{eq.deltax}) is the least square solution, which might be incorrect when $\bar{\boldsymbol{H}}^{\mathrm{T}}\bar{\boldsymbol{H}}$ is approximately ill-conditioned. In this case, the classical least square problem can be converted to be a regularized least square problem by exploiting the regularization on $\delta\boldsymbol{x}(t_{0}^{j})$. The detailed discussion can be found in Section \ref{sect.framwork}.

When $\tilde{\boldsymbol{x}}(t_{0}^{j})+\hat{\delta\boldsymbol{x}}(t_{0}^{j})$ is substituted into $\tilde{\boldsymbol{x}}(t_{0}^{j})$, we can start a new round of iteration to estimate $\hat{\delta\boldsymbol{x}}(t_{0}^{j})$. The improved algorithm is summarized as \textbf{Algorithm 2}.

\begin{table}[h]\label{algorithm2}
  \begin{tabular}{l}
  \hline
      \textbf{Algorithm 2} Iterated Estimation of $\phi_{0}$, $\bar{\nu}_{1}$, $\bar{\nu}_{2}$ and $\boldsymbol{x}(t_{0}^{j})$\\ \hline
      1: \textbf{Initialization}: \\
      ~~~~Divide the $j$th exposure into $M$ sub-exposures;\\
      ~~~~Set $\bar{\boldsymbol{\nu}}^{(0)}=\left[\phi_{0}^{(0)}, \bar{\nu}_{1}^{(0)}, \bar{\nu}_{2}^{(0)}\right]^{\mathrm{T}}$ and $\tilde{\boldsymbol{x}}(t_{0}^{j})=\left[\tilde{\boldsymbol{r}}^{\mathrm{T}}(t_{0}^{j}), \tilde{\boldsymbol{v}}^{\mathrm{T}}(t_{0}^{j})\right]^{\mathrm{T}}$\\
      2: \textbf{for} $q=1,\cdots, Q$ \textbf{do}\\
      3: ~~~~Apply \textbf{Algorithm 1} to get $\hat{\bar{\boldsymbol{\nu}}}$;\\
      4: ~~~~Re-calculate $\phi$ based on $\hat{\bar{\boldsymbol{\nu}}}$, (\ref{phapropag_ploy}), and GEF;\\
      5: ~~~~Estimate $\hat{\delta\boldsymbol{x}}(t_{0}^{j})$ based on (\ref{eq.BOD1})-(\ref{eq.deltax});\\
      6: ~~~~\textbf{if} $\|\hat{\delta\boldsymbol{x}}(t_{0}^{j})\|<\epsilon $\\
      7: ~~~~~~~~break;\\
      8:~~~~\textbf{else}\\
      11:~~~~~~~~$\tilde{\boldsymbol{x}}(t_{0}^{j})\leftarrow\tilde{\boldsymbol{x}}(t_{0}^{j})+\hat{\delta\boldsymbol{x}}(t_{0}^{j})$;\\
      12:~~~~~~~~$q\leftarrow q+1$;\\
      13:~~~~\textbf{end if}\\
      14:\textbf{end for}\\
      15:\textbf{Output}: $\hat{\bar{\boldsymbol{\nu}}}$ and $\hat{\boldsymbol{x}}(t_{0}^{j})$.\\ 
      \hline
  \end{tabular}
\end{table}

The computational complexity of \textbf{Algorithm 2} is about $O\left((K+1)M^{3}Q+KMN_{b}^{2}Q\right)$. When it is applied to the same example in Section \ref{subsect.mot} with $Q$ as 3, the computational complexity of \textbf{Algorithm 2} is about $O(5.4\times 10^{6})$. 

For comparison, we provide the procedure of Significance Enhancement of Pulse-profile with Orbit-dynamics (SEPO) as the \textbf{Algorithm for Comparison 2}. SEPO was proposed in \cite{Zheng_2019} to estimate the orbit elements of a satellite at the initial time of an exposure. Given that the orbit elements can be transformed to be position and velocity, we employ the SEPO to estimate $\boldsymbol{x}(t_{0}^{j})=\left[x, y, z, v_{x}, v_{y}, v_{z}\right]^{\mathrm{T}}$. If $N_{x}$, $N_{y}$, $N_{z}$, $N_{v_{x}}$, $N_{v_{y}}$, $N_{v_{z}}$ are all set as 1000, in the same example in Section \ref{subsect.mot}, the computational complexity of SEPO is about $O(10^{18})$. Thus, the computational complexity of \textbf{Algorithm 2} is about $10^{-12}$ of SEPO. 

\begin{table}[h]\label{algorithm4}
  \begin{tabular}{l}
  \hline
      \textbf{Algorithm for Comparison 2} 
      ~SEPO for estimating $\boldsymbol{x}(t_{0}^{j})$\\
       \hline
      1: \textbf{Initialization}:\\
      2: ~~~~Assume the search spaces for $x, y, z, v_{x}, v_{y}, v_{z}$ are \\
      ~~~~~~~~$[x_{\mathrm{min}}, x_{\mathrm{max}}]$, $[y_{\mathrm{min}}, y_{\mathrm{max}}]$,
      $[z_{\mathrm{min}}, z_{\mathrm{max}}]$, $[v_{x,\mathrm{min}}, v_{x,\mathrm{max}}]$\\
      ~~~~~~~~$[v_{y, \mathrm{min}}, v_{y, \mathrm{max}}]$, and $[v_{z,\mathrm{min}}, v_{z, \mathrm{max}}]$\\
      3:~~~~Divde The search spaces for $x, y, z, v_{x}, v_{y}, v_{z}$ into $N_{x}$, $N_{y}$,\\
      ~~~~~~~~$N_{z}$, $N_{v_{x}}$, $N_{v_{y}}$, $N_{v_{z}}$ segments respectively.\\ 
      4: ~~~~Design a $N_{x} \times N_{y} \times N_{z} \times N_{v_{x}} \times N_{v_{y}}\times N_{v_{z}}$ grid;\\
      5: \textbf{for} $k=1,\cdots, N_{x}$ \textbf{do}\\
      6: ~~~~$x^{(k)}=x_{\mathrm{min}}+\frac{k-1}{N_{x}}\left(x_{\mathrm{max}}-x_{\mathrm{min}}\right)$\\
      7: ~~~~\textbf{for} $l=1,\cdots, N_{y}$ \textbf{do}\\
      8: ~~~~~~~~$y^{(l)}=y_{\mathrm{min}}+\frac{l-1}{N_{y}}\left(y_{\mathrm{max}}-y_{\mathrm{min}}\right)$\\
      9: ~~~~~~~~\textbf{for} $u=1,\cdots, N_{z}$ \textbf{do}\\
      10: ~~~~~~~~~~~~~~$z^{(u)}=z_{\mathrm{min}}+\frac{u-1}{N_{z}}\left(z_{\mathrm{max}}-z_{\mathrm{min}}\right)$\\
      11: ~~~~~~~~~~~~~~\textbf{for} $b=1,\cdots, N_{v_{x}}$ \textbf{do}\\
      12: ~~~~~~~~~~~~~~~~~~~$v_{x}^{(b)}=v_{x, \mathrm{min}}+\frac{b-1}{N_{v_{x}}}\left(v_{x, \mathrm{max}}-v_{x, \mathrm{min}}\right)$\\
      13: ~~~~~~~~~~~~~~~~~~~\textbf{for} $a=1,\cdots, N_{v_{y}}$ \textbf{do}\\
      14: ~~~~~~~~~~~~~~~~~~~~~~~$v_{y}^{(a)}=v_{y, \mathrm{min}}+\frac{v-1}{N_{v_{y}}}\left(v_{y, \mathrm{max}}-v_{y, \mathrm{min}}\right)$\\
      15: ~~~~~~~~~~~~~~~~~~~~~~~\textbf{for} $p=1,\cdots, N_{v_{z}}$ \textbf{do}\\
      16: ~~~~~~~~~~~~~~~~~~~~~~~~~~~$v_{z}^{(p)}=v_{z, \mathrm{min}}+\frac{p-1}{N_{v_{z}}}\left(v_{z,\mathrm{max}}-v_{z,\mathrm{min}}\right)$\\
      17: ~~~~~~~~~~~~~~~~~~~~~~~~~~~Propagate an orbit through the $j$th\\ ~~~~~~~~~~~~~~~~~~~~~~~~~~~~~exposure initialized with $\boldsymbol{x}=
      \left[\begin{aligned}
        x^{(k)}, y^{(l)}, z^{(u)}, \\
        v_{x}^{(b)}, v_{y}^{(v)}, v_{z}^{(p)}
      \end{aligned}
      \right]^{\mathrm{T}}$\\
      ~~~~~~~~~~~~~~~~~~~~~~~~~~~~~and calculate the significance of the pulse profile\\
        ~~~~~~~~~~~~~~~~~~~~~~~~~~~~~$\chi^{2}(x^{(k)}, y^{(l)}, z^{(u)}, v_{x}^{(b)}, v_{y}^{(v)}, v_{z}^{(p)})$ which is \\
        ~~~~~~~~~~~~~~~~~~~~~~~~~~~~~ defined in (1) in \cite{Zheng_2019}\\
      18: ~~~~~~~~~~~~~~~~~~~~~~~\textbf{end for}\\
      19: ~~~~~~~~~~~~~~~~~~~\textbf{end for}\\
      20: ~~~~~~~~~~~~~~~\textbf{end for}\\
      21: ~~~~~~~~~~~~\textbf{end for}\\
      22: ~~~~~~~~\textbf{end for}\\
      23: ~~~~\textbf{end for}\\
      24:\textbf{end for}\\
      25: $\hat{x}, \hat{y}, \hat{z}, \hat{v_{x}},\hat{v_{y}}, \hat{v_{z}}= \arg \max \chi^{2}$\\
      26:\textbf{Output:} $\hat{\boldsymbol{x}}(t_{0}^{j})=\left[\hat{x}, \hat{y}, \hat{z}, \hat{v_{x}},\hat{v_{y}}, \hat{v_{z}}\right]^{\mathrm{T}}$\\ 
      \hline
  \end{tabular}
\end{table}

\section{Experiments and Results}
\label{Sect.ER}   
In this section, we employ the Crab pulsar data from \textit{Insight}-HXMT and \textit{NICER} to verify the proposed algorithm.

\subsection{Description of Data}
\subsubsection{Data description for \textit{Insight}-HXMT} The experiment utilizes two data sets. The first set was acquired over the period from 2018 October 30th through 2018 November 1st (ObsID: P0101299008), and the second set was obtained between 2017 August 31st and the September 2nd (ObsID: P0101299002). The data reduction is performed according to the criteria proposed in \cite{Zheng_2019}. In the navigation experiment, the initial position and velocity of \textit{Insight}-HXMT is set as the Global Positioning System (GPS) solution with a (8 km, 8 km, 8 km, 5 m/s, 5 m/s, 5 m/s) Earth-centered error.

\subsubsection{Data description for \textit{NICER}} The data of \textit{NICER} on the 2018 December 26th (ObsID: 1013010147) is employed. The criteria for data reduction is employed according to \cite{Deneva_2019}. As a result, there are 12 exposures. The state of ISS is initialized by the state provided by the Heasoft v.26.1 with a (15 km, 15 km, 15 km, 2 m/s, 2 m/s, 2 m/s) Earth-centered error.

\subsection{Results}
In this section, XTITAN refers to \textbf{Algorithm 2} shown in Section IV. Regarding that the purpose of pulse phase estimation is to estimate the position and velocity of satellite, we investigate the estimation performance of XTITAN by assessing the root mean square error (RMSE) of the estimated position and velocity relative to the position and velocity provided by GPS or by the Heasoft v.26.1.

\begin{figure}[htbp]
    \centering
    \includegraphics[width=\columnwidth]{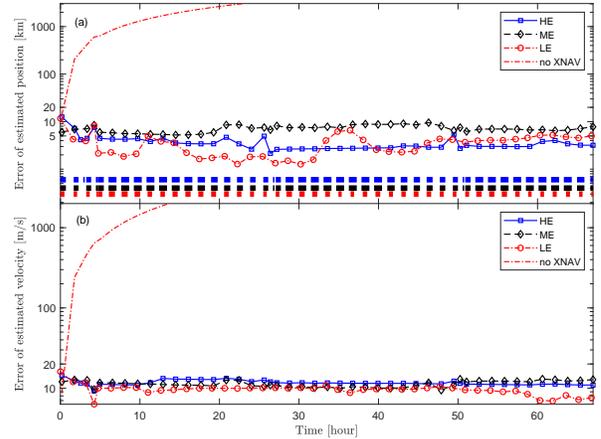}
    \caption{Estimation result of the proposed method using 2018 \textit{Insight}-HXMT data: (a) the error of estimated position and (b) the error of estimated velocity.}
    \label{fig:result_HXMT_2018_3}
\end{figure}

XTITAN is first sequentially applied to the data from \textit{Insight}-HXMT in 2018. Figure \ref{fig:result_HXMT_2018_3} shows the position and velocity estimation results. The blue, black, and red bars in the Figure \ref{fig:result_HXMT_2018_3}.(a) present the exposures on pulsar from the High Energy detector (HE), the Middle Energy detector (ME), and the Low Energy detector (LE) respectively. As shown in Figure \ref{fig:result_HXMT_2018_3}.(a), there are gaps between two consecutive exposures. The reasons for the gaps include that the pulsar was occulted by the Earth and that the data was reduced according to the data reduction criteria. The exposures and gaps vary with time and with detectors because that the space environment varies with time and that the background noise of detectors are different. The data from the three detectors onboard \textit{Insight}-HXMT can all ensure the convergence of error of estimated position and velocity. Although the estimated errors for the three detectors present slightly different trends, most of them are about 5 km. In contrast, if there was no pulsar observed, the position error rapidly grows as time increases.

\begin{figure}[htbp]
  \centering
  \includegraphics[width=\columnwidth]{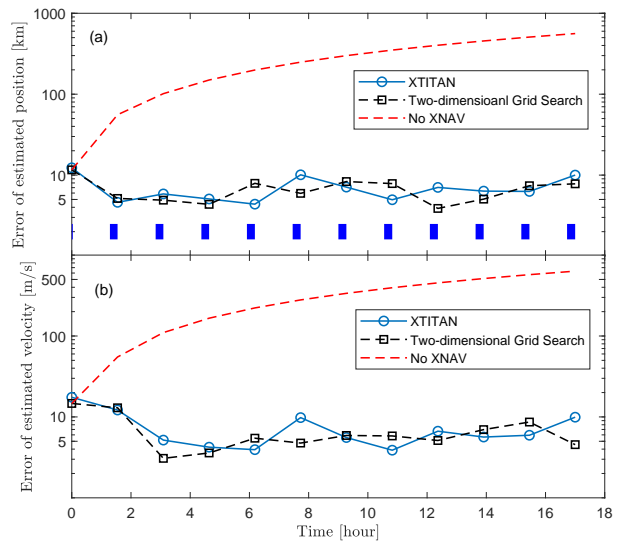}
  \caption{Performance comparison between XTITAN and the two-dimensional grid search using \textit{NICER} data : (a) the error of estimated position and (b) the error of estimated velocity.}
  \label{fig:result_NICER}
\end{figure}

When XTITAN is applied to the data from \textit{NICER}, Figure \ref{fig:result_NICER} shows the position and velocity estimation results obtained from XTITAN and the two-dimensional grid search. The blue bars in Figure \ref{fig:result_NICER}.(a) indicate the 12 exposures on pulsar. Compared with the data of \textit{Insight}-HXMT, the exposures and gaps of \textit{NICER} data distribute more evenly. The durations of the exposures are around 2000 s, and the gaps are around 3000 s. As time increases, the error of estimated position converges to around 5 km when XTITAN or the two-dimensional grid search is applied. By contrast, the estimated error will dramatically grow if where are not exposures on pulsar. In addition, the estimated error curves for XTITAN and for two-dimensional grid search are close to each other. Compared with Figure \ref{fig:result_HXMT_2018_3}, the estimation error curves for \textit{NICER} are more steady than \textit{Insight}-HXMT. It is because that the exposures of \textit{NICER} are all about 2000 s over the whole navigation process and then the pulse phase estimations at each exposure have similar accuracies.

\begin{figure}[htbp]
    \centering
    \includegraphics[width=\columnwidth]{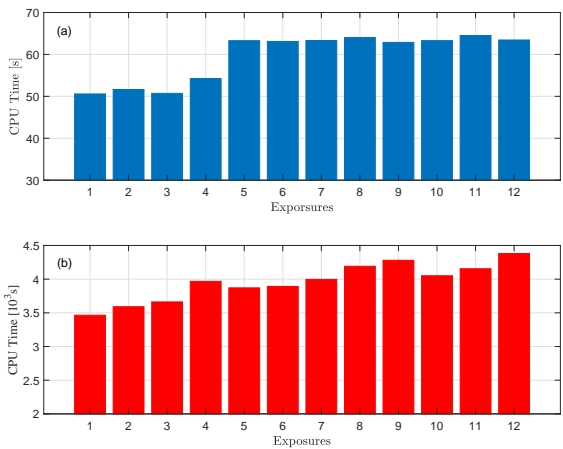}
    \caption{CPU time cost by XTITAN and by the two-dimensional grid search: (a) CPU time cost by XTITAN and (b) CPU time cost by the two-dimensional grid search.}
    \label{fig:ET_NICER}
\end{figure}

In the computation environment including the Intel Core i7-4790 CPU @3.6GHz and the python 3.75, Figure \ref{fig:ET_NICER} shows the CPU time cost by XTITAN and by the two-dimensional grid search. The CPU times of XTITAN are around 60 s, but the CPU times of the two-dimensional grid search are all above 3500 s. Moreover, the CPU times for the first three exposures are all less than the other exposures because that the amount of events in the three exposures are less than the other exposures. In practice, the pulse phase estimation starts when an exposure accomplished. Given that the gaps between two exposures of \textit{NICER} data are around 3000 s, the two-dimensional grid search cannot finish computing before a new exposure starts. Then, it will cause a disaster to the navigation process. In contrast, XTITAN is about 58 times faster than the two-dimensional grid search, and its CPU time is much less than the gap. Thus, XTITAN is more suitable for the onboard computation of Crab pulsar data than the two-dimensional grid search. 

\begin{figure}[htbp]
    \centering
    \includegraphics[width=\columnwidth]{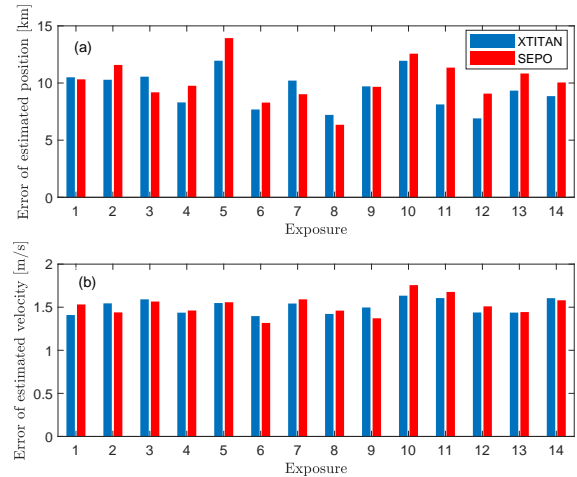}
    \caption{Comparison between XTITAN and SEPO: (a) the error of estimated position and (b) the error of estimated velocity.}
    \label{fig:SEPO_Comparison}
\end{figure}

We further compare XTITAN with SEPO. Given that the SEPO could only estimate the position and velocity of satellite at the initial time of an exposure, we investigate the position and velocity estimation performance at the initial times of 14 3000s-exposures of \textit{Insight}-HXMT obtained in 2017, which is shown in Table \ref{tab:True state}. As shown in Figure \ref{fig:SEPO_Comparison}, XTITAN can provide estimation errors smaller than SEPO. In addition, the computational time of SEPO is about 3 hours, but XTITAN only takes 60 s in the same computation environment. Thus, XTITAN is much computationally efficient than SEPO.

\begin{table}
  \footnotesize
  \caption {Exposures for comparison between XTITAN and SEPO}
  \scriptsize{}
  \label{tab:True state}
  \begin{center}
  \begin{tabular}{clcl}
  \hline
  No. & Start and Finish Time [UTC] & No. & Start and Finish Time [UTC]\\
  \hline
  1 & 2017.8.31	13:17:20-14:07:20 & 8 & 2017.9.01	00:30:26-01:20:26\\
  2 & 2017.8.31	14:52:46-15:42:46 & 9 & 2017.9.01	02:02:06-02:52:06\\
  3 & 2017.8.31	16:28:12-17:18:06 & 10 & 2017.9.01 03:42:06-04:32:06\\
  4 & 2017.8.31	18:07:06-18:57:06 & 11 & 2017.9.01 05:13:46-06:03:46\\
  5 & 2017.8.31	19:39:05-20:29:05 & 12 & 2017.9.01 16:20:26-17:10:26\\
  6 & 2017.8.31	21:18:46-22:08:46 & 13 & 2017.9.02 20:57:51-21:47:51\\
  7 & 2017.8.31	22:50:26-23:40:26 & 14 & 2017.9.02 22:37:06-23:27:06\\
  \hline
  \end{tabular}
  \end{center}
\end{table}

\section{Conclusion}
In this paper, we propose an X-ray pulsar-based navigation using on-orbit pulsar timing (XTITAN). At each exposure, XTITAN first approximates the pulse phase evolution with the aid of orbit dynamics of satellite, and corrects the approximation by performing an on-orbit pulsar timing instead of the grid search. XTITAN is improved to iteratively estimate the position and velocity of satellite at the start time of the exposure as well as to correct the pulse phase propagation approximation. When applied to the Crab pulsar data from \textit{NICER}, XTITAN is 58 times faster than the two-dimensional grid search. When applied to the Crab pulsar data from \textit{Insight}-HXMT, XTITAN is 180 times faster than the Significance Enhancement of Pulse-profile with Orbit-dynamics (SEPO) which was employed in the flight experiments on \textit{Insight}-HXMT.


%

\section*{Acknowledgment}
This work is funded by The National Natural Science Foundation of China (No. 61703413) and the Science and Technology Innovation Program of Hunan Province (No. 2021RC3078).


\ifCLASSOPTIONcaptionsoff
  \newpage
\fi



\bibliographystyle{IEEEtran}
\bibliography{XNAV.bib}
%
%

%








\end{document}